\documentclass[prc,twocolumn,epsfig,nofootinbib]{revtex4}
\usepackage{graphics}
\usepackage{epsfig}
\usepackage{amsfonts}
\usepackage{amsmath}
\usepackage{bm}

\usepackage{color}

\begin{document}
\title{Deformation-induced splitting of the monopole giant resonance in $^{24}$Mg}

\author{J. Kvasil$^{1}$, V.O. Nesterenko$^{2}$, A. Repko$^{1}$, P.-G. Reinhard$^{3}$ and  W. Kleinig$^{2,4}$}
\affiliation{$^1$
 Institute of Particle and Nuclear Physics, Charles University, CZ-18000, Praha 8, Czech Republic}
\affiliation{$^2$
Laboratory of Theoretical Physics, Joint Institute for Nuclear Research, Dubna, Moscow region, 141980, Russia}
\affiliation{$^3$
Institut f\"ur Theoretische Physik II, Universit\"at Erlangen, D-91058, Erlangen, Germany}
\affiliation{$^4$
Technische Universit\"at Dresden, Institut f\"ur Analysis, D-01062, Dresden, Germany}

\date{\today}

\begin{abstract}
The strong deformation splitting of the isoscalar giant monopole resonance (ISGMR),
recently observed in ($\alpha,\alpha'$) reaction in prolate $^{24}$Mg, is analyzed
in the framework of the Skyrme quasiparticle random-phase-approximation (QRPA)
approach with the Skyrme forces SkM*, SVbas and SkP$^{\delta}$.
The calculations with these forces
give close results and confirm that the low-energy E0-peak is caused by the
deformation-induced coupling of ISGMR with the K=0 branch of the isoscalar giant
quadrupole resonance.
\end{abstract}
\maketitle
\section{Introduction}
\label{intro}

The isoscalar giant monopole  resonance (ISGMR) represents the main source of
information on the nuclear incompressibility \cite{Bl80} and is a subject of
intense studies during last decades, see the book \cite{Ha01}
and recent reviews \cite{Colo08,Av13}. In deformed nuclei, ISGMR is
coupled with the $K^{\pi}=0^+$ branch of the isoscalar  giant quadrupole
resonance (ISGQR), which leads to a double-bump structure (splitting) of the
ISGMR \cite{Ha01}. The energy interval between the bumps is larger than the
ISGMR and ISGQR widths \cite{Ha01}. Besides, in well deformed nuclei, both bumps carry a
significant monopole strength. These two factors favor an experimental
observation of the ISGMR splitting.

In medium and heavy deformed nuclei, the clear  ISGMR
splitting has been found only in Sm isotopes
\cite{Garg_80,Young_99,Young_04,Itoh_03} and $^{238}$U\cite{Bra_82}.
In this aspect, light deformed nuclei look more promising because some of them demonstrate
much stronger deformation ($\beta$=0.5-0.60)\cite{Raman} than well deformed heavier nuclei
($\beta$=0.30-0.35). Thus we may expect in light nuclei particularly strong E0-E2 coupling.

The first experimental data of this kind have been recently obtained.
Namely, the ISGMR splitting in $^{24}$Mg  was observed in $(\alpha,\alpha')$
reaction to forward angles \cite{Gupta_15}. The experiment was performed in the
Research Center for Nuclear Physics (RCNP) in Osaka University. The nucleus $^{24}$Mg
has a huge prolate quadrupole axial deformation with $\beta$=0.605$\pm$0.008 \cite{Raman}
and thus promises a strong E0-E2 coupling. However fist attempts to observe a discernible
ISGMR splitting in this nucleus using inelastic scattering of $\alpha$-particles and $^{6}$Li
have failed \cite{Young_99,Chen_09,Young_09,Kaw_13}.
Only in \cite{Gupta_15}, the experimentalists
have managed to reliably  discriminate the splitting. It turned out
to be huge, with a strong narrow peak at
$E_1\sim$ 16 MeV and a broad structure at $E_2\sim$ 24 MeV, see Fig. 1.

The calculations in the framework of the quasiparticle random-phase-approximation (QRPA)
with the Skyrme force SkM* \cite{SkMs}, presented in Ref. \cite{Gupta_15}, have confirmed the E0-E2
origin of the ISGMR splitting observed in $^{24}$Mg. However it is known that
QRPA description of giant resonances can noticeably depend on the applied Skyrme parametrization
\cite{Ben03,Nes_IJMPE_07,Nes_IJMPE_08,SV}. Such dependence can take place for
ISGMR in deformed nuclei as well \cite{Kva_15}. In this connection, it is worth
to check the Skyrme QRPA results obtained with SkM* in Ref. \cite{Gupta_15}
by using other Skyrme parameterizations. This is just the aim of the present study.

Here we explore the ISGMR splitting in $^{24}$Mg within Skyrme QRPA
approach with the forces SkM* \cite{SkMs}, SVbas \cite{SV} and SkP$^{\delta}$
\cite{SkPd}. The force SkM* (K$_{\infty}$=217 MeV) is used
for comparison with the previous calculations \cite{Gupta_15}. Two other forces
are chosen as representatives of essentially different nuclear
incompressibilities: K$_{\infty}$=234 MeV  for SVbas
and 202 MeV  for SkP$^{\delta}$. As shown below, all three forces give qualitatively
close results and confirm that the ISGMR splitting arises just because of the
E0-E2 coupling.
\begin{figure*}
\centering \resizebox{0.6\textwidth}{10cm} {\includegraphics {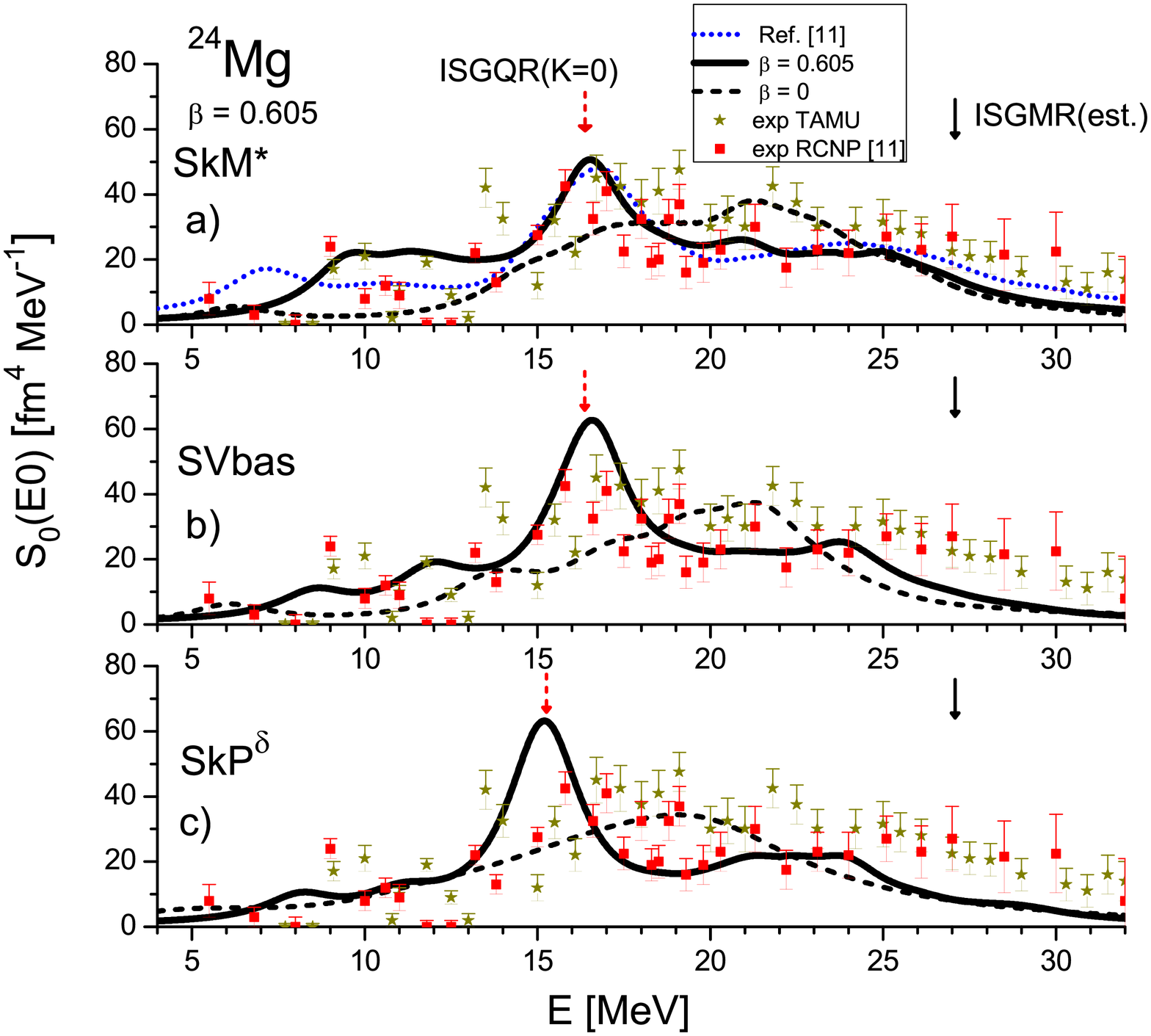}}
\caption{The QRPA E0(T=0) strength functions in $^{24}$Mg, calculated  at the
experimental quadrupole deformation $\beta$=0.605 (black solid curves) and
in the spherical limit $\beta$=0 (black dash curves) with the Skyrme forces SkM*
(a), SVbas (b), and SkP$^{\delta}$ (c). In the panel a), the SkM* strength function
from \cite{Gupta_15} is also given (blue dotted curve). In all the panels, the
TAMU \cite{Gupta_15,TAMU} (yellow stars) and RCNP \cite{Gupta_15} (red squares) experimental data are
exhibited. The calculated energy centroids of the ISGQR(K=0)-branch  are
indicated by the red dash arrows. The estimation $E_{\rm{ISGMR}}= 78 A^{-1/3}$ MeV \cite{Ha01}
for the ISGMR energy is marked by the black bold arrow.} \label{fig-1}
\end{figure*}

\section{Model}

The calculations are  performed with the two-dimensional (2D) QRPA code
\cite{Repko}. As compared to our previous calculations for ISGMR \cite{Kva_15},
the present code does not use any separable ansatz.
The method is fully self-consistent because: i) both the mean field and
residual interaction are obtained from the same Skyrme functional, ii)
the residual interaction includes all terms of the functional,
including Coulomb (direct and exchange) terms. Both time-even
and time-odd densities are involved. The code exploits a mesh in
cylindrical coordinates with a mesh size of d=0.7 fm and a calculation
box of about three nuclear radii.

The experimental value  of the deformation $\beta$=0.605 \cite{Raman} is used for all three forces.
The $\delta$-force volume pairing  is treated at the BCS level
\cite{Ben00}. The pairing particle-particle channel in the residual interaction
is taken into account.

 The ISGMR strength function reads
\begin{equation}
S(E0; E) = \sum_{\nu} |\:\langle \nu | \hat{M}(E0) | 0
\:\rangle |^2 \: \xi_{\Delta}(E-E_\nu ) \label{4}
\end{equation}
where $|0\rangle$ is the ground state wave function, $|\nu\rangle$ and
$E_{\nu}$ are QRPA states and energies,
$\hat{M}_{\rm{ISGMR}}(E0)=\sum_i^A (r^2Y_{00})_i$ is the isoscalar (T=0)
transition operator, $\xi_{\Delta}(E-E_{\nu}) = \Delta /(2\pi
[(E-E_{\nu})^2 - \Delta^2/4]$ is the Lorentz smoothing with the averaging
parameter $\Delta$. The Lorentz function approximately  simulates smoothing effects
beyond QRPA and makes convenient comparison of the calculated and
experimental strengths. The calculations \cite{Gupta_15}used
the smearing 3 MeV. Here the averaging $\Delta$= 2.5 MeV is found optimal.

Our calculations use sufficiently large basis. In SVbas the single
particle (s-p) spectrum includes 750 proton and 920 neutron levels
in the energy intervals (-32, +61) MeV
and (-37, +85) MeV, respectively. The two-quasiparticle basis involves about 3700 states
with the energies up to $\sim$200 MeV. The energy-weighted sum rule for the
isoscalar monopole excitations is exhausted by 94$\%$. The similar basis is used
for SkM* and SkP$^{\delta}$.

\section{Results and discussion}

Results of the calculations are presented in Fig. 1. They are compared
with the experimental  data from RCNP \cite{Gupta_15} and TAMU (Texas
A$\&$M University) \cite{Gupta_15,TAMU}. Following the statement in
Ref. \cite{Gupta_15}, RCNP data do exhibit the
ISGMR splitting  while TAMU data do not.

As seen from Fig. 1, the calculations with SkM*, SVbas and SkP$^{\delta}$ give
a qualitatively similar picture and justify origin of the narrow peak at
$E\sim$ 16 MeV as a result of  the coupling between ISGMR and K=0 branch of
ISGQR. Indeed the position of this peak well coincides with the energy of
ISGQR(K=0) branch, marked by the dash arrows in the figure. The second evidence
is that this peak is almost disappears in the spherical limit ($\beta$=0) when
the E0-E2 coupling is absent. The agreement with the experimental energy
$E\sim$ 16 MeV is very nice for SkM* and SVbas and somewhat worse
for SkP$^{\delta}$. The latter is explained by a
low incompressibility  K$_{\infty}$=202 MeV in SkP$^{\delta}$.

Our and previous \cite{Gupta_15} SkM* calculations give about the same
description of the peak at $E\sim$ 16 MeV (compare the black bold and blue dotted
curves in the panel (a)). Both results are also similar for E0 strength
above the peak. A noticeable difference between two calculations takes place only
in a low energy region with E$<$13 MeV, which is beyond of our interest. Perhaps
this difference is partly caused by using
variant pairing schemes: (HF+BCS in our case
and HFB in \cite{Gupta_15}).

In Figure 1, the RCNP experimental data give a high-energy distribution peaked at $\sim$24 MeV.
This distribution constitutes the familiar ISGMR. The broad ISGMR structure seems to continue
to the low-energy region and form a massive background of the narrow peak at $E\sim$ 16 MeV.
This is confirmed by the distribution of E0 strength calculated in the spherical limit.
Note that the energy $E\sim$ 24 MeV gives $E_{\rm{ISGMR}}= 69 A^{-1/3}$ MeV which is closer
to the empirical estimation for ISGQR ($E_{\rm{ISGQR}}= 64 A^{-1/3}$ MeV) rather
than for ISGMR ($E_{\rm{ISGMR}}= 78 A^{-1/3}$ MeV) \cite{Ha01}. The latter
gives for ISGMR in $^{24}$Mg the energy $E_{\rm{ISGMR}}\sim$27 MeV.

As seen from Fig. 1, our calculations in general reproduce the ISGMR at
$\sim$24 MeV. The forces SkM* and SVbas demonstrate a better performance than
SkP$^{\delta}$ with its very low incompressibility.
Note also that present SkM*, SVbas and SkP$^{\delta}$ calculations,
as well as the SkM* results \cite{Gupta_15}, underestimate the observed E0 strength
at E$>$ 24 MeV.
Thus the computed ISGMR looks somewhat downshifted as compared to the
experimental data.
Most probably the QRPA is not enough to describe details of present experimental
distribution  and we need here the coupling with complex configurations.

\section{Summary}

The analysis of the recent experimental data for the deformation-induced splitting of
the isoscalar giant monopole resonance (ISGMR) in strongly deformed $^{24}$Mg
\cite{Gupta_15} has been done within the Skyrme quasiparticle random-phase
approximation (QRPA) approach. The Skyrme forces SkM* \cite{SkMs}, SVbas \cite{SV},
and SkP$^{\delta}$ with different values of nuclear incompressibility were applied.
The calculation generally well reproduce the experimental data and confirmed
the origin of the peak at $E\sim$ 16 MeV as a result of the E0-E2 coupling. For a more
precise description of the experimental E0 distribution, inclusion of the coupling
with complex configurations is desirable.

\section*{Acknowledgments}
The work was partly supported by the DFG grant RE 322/14-1, Heisenberg-Landau
(Germany-BLTP JINR), and Votruba-Blokhintsev (Czech Republic-BLTP JINR)
grants. The BMBF support under the contracts 05P12RFFTG (P.-G.R.) and 05P12ODDUE
(W.K.) is appreciated. J.K. is grateful for the support of the Czech Science
Foundation (P203-13-07117S). We thank U. Garg for useful discussions.

\end{document}